\documentclass[aps,prl,twocolumn,showpacs,superscriptaddress,groupedaddress]{revtex4}  
\usepackage{amssymb,amsmath,graphicx}
\usepackage{graphicx,wrapfig,lipsum}
\usepackage[usenames,dvipsnames]{xcolor}
\usepackage{graphicx}  
\usepackage{dcolumn}   
\usepackage{bm}        
\usepackage{amssymb}   
\begin{document}
\title{Kepler orbits of settling discs}
\author{Rahul Chajwa$^*$}
\affiliation{International Centre for Theoretical Sciences, Tata Institute of Fundamental Research, Bengaluru 560 089}
\author{Narayanan Menon$^*$}
\affiliation{Department of Physics, University of Massachusetts, Amherst MA 01003 USA}
\author{Sriram Ramaswamy$^*$}
\affiliation{Department of Physics, Indian Institute of Science, Bengaluru 560 012\\$^*$TIFR Centre for Interdisciplinary Sciences, Hyderabad 500 107} 


\begin{abstract}

The collective dynamics of objects moving through a viscous fluid is complex and counterintuitive \cite{HB1,kim1,Chaikin,SR1,anderson,bradybossis,lauga}. A key to understanding the role of nontrivial particle shape in this complexity is the interaction of a pair of sedimenting spheroids  \cite{jeffery1,wakiya,chwang,kim2,shelley,witten1}. We report experimental results on two discs settling at negligible Reynolds number ($\simeq 10^{-4}$), finding two classes of bound periodic orbits, each with transitions to scattering states. We account for these dynamics, at leading far-field order, through an effective Hamiltonian in which gravitational driving endows orientation with the properties of momentum. This leads to a precise correspondence with the Kepler problem of planetary motion for a wide range of initial conditions, and also to orbits with no Keplerian analogue. This notion of internal degrees of freedom manifesting themselves as an effective inertia is potentially a more general tool in Stokesian driven systems.
\end{abstract}
\maketitle

Particles settling in a fluid carry monopoles of force density \cite{Stokes1851}. In the 
Stokesian limit of Reynolds number Re $\to 0$ they therefore manifest the hydrodynamic interaction in its strongest form. Among the consequences of this strong coupling are chaos in three-particle settling \cite{janosi, Ekiel}, and the resulting statistical character of many-particle sedimentation \cite{SR1,Guazzelli,witten2, Vijay}. Interestingly, however, the collective settling of identical spheres can be built up from two-particle processes \cite{jeffery2,crowley,LR}: i) a pair falls faster than an isolated sphere, with a horizontal drift when their separation is oblique to gravity. The reversibility of Stokes flow \cite{HB1} ensures that the separation vector stays constant. By the same token a single apolar axisymmetric particle does not rotate, and drifts horizontally as it falls. Two non-spherical particles display a far richer sedimentation dynamics, via mutual rotation
due to a coupling between orientational and translational degrees of freedom \cite{witten1,shelley}. In this Letter, we present experiments that classify the possible dynamical
behaviours of a settling pair of discs. We show that a symmetry-based
far-field theory, without a detailed calculation of the mutual rotation
coupling, accounts for the dynamics through the emergence of an effective
Hamiltonian for this wholly dissipative system.

\begin{figure}[t]
\begin{center}
\includegraphics[width=8.5cm]{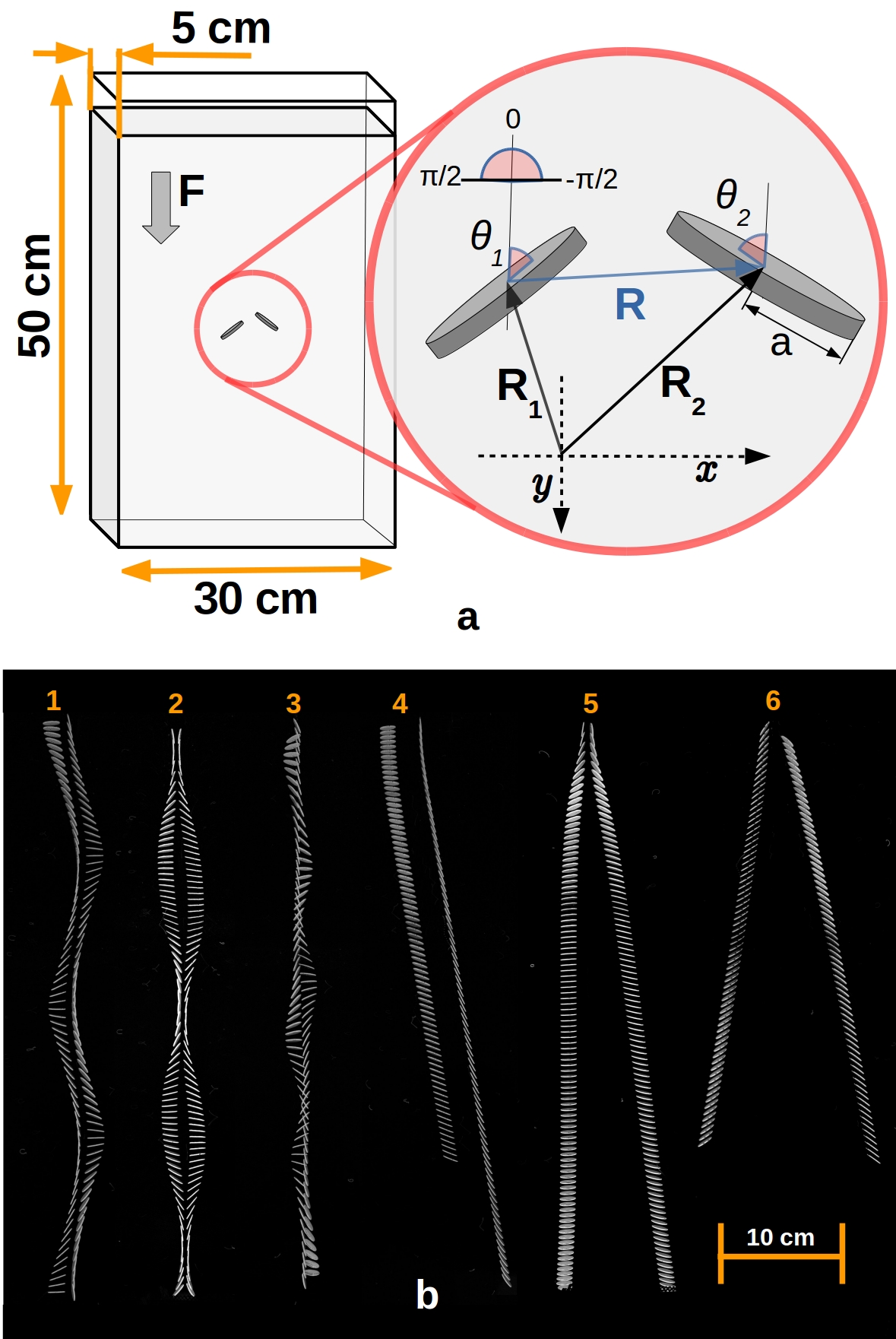}
\caption{\label{fig:setup}\textbf{Bound and scattering behaviour}: (a) A quasi-two-dimensional setup with discs released such that the vector normal to the disc and the separation vector $ \vec{R} = \vec{R}_{2} - \vec{R}_{1} $ lie in the plane of the settling geometry, (x,y). The orientation of individual disc is quantified by angles $\theta_{1}$ and $\theta_{2}$ measured w.r.t gravity pointing along the $\hat{y}$ direction. (b) The z-stacks of overlapped time frames showing pair dynamics observed in experiments. The dynamics are generated by varying the initial separation ($x_{o}$, $y_{o}$) between the discs and their individual orientations ($\theta_{i}; i=1,2$). These complex trajectories can be grouped into two broad classes: periodic bound (1-3) and scattering (4-6).}
\end{center}
\end{figure}

Our experiments are conducted on pairs of identical discs, with radius $a=0.65$ cm, falling in viscous fluid (Re $\sim 10^{-4}$) (see Methods). As shown in Figure \ref{fig:setup}(a) the trajectory of the centres of the discs lie in a plane. Assuming translation symmetry in this plane, the six coupled degrees of freedom can be reduced to two separation and two orientation degrees of freedom. Our observations suggest two qualitatively distinct trajectory types: scattering, in which the separation increases monotonically, and bound, in which separation and orientations oscillate with a characteristic amplitude and wavelength. The oscillatory behaviour further falls into two classes, to be discussed later. 

We ask: (i) Is there a well-defined boundary in the space of initial conditions that separates periodic and scattering (i.e. infinite-wavelength) behaviour, or do our ``scattering'' states simply have a wavelength longer than the container height? (ii) What determines the emergent time period and wavelength of the periodic orbits?  

\begin{figure}[h]
\begin{center}
\includegraphics[width=8.5cm]{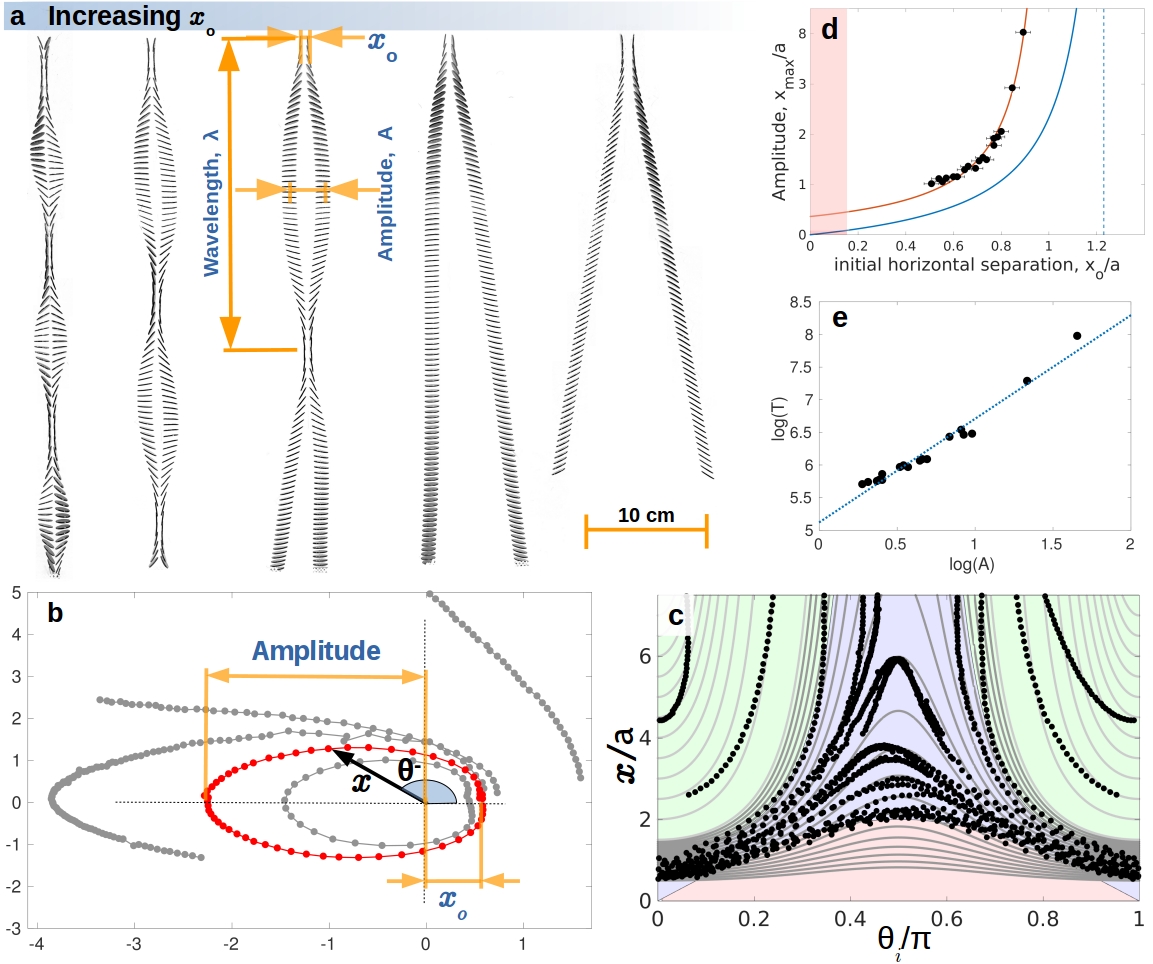}
\caption{\label{fig:symmetric}\textbf{Symmetric settling}: (a) Overlapped time-lapse images from the experiment, exhibiting a transition from periodic orbits to scattering trajectories with increasing initial horizontal separation $x_{o}$. Wavelength $\lambda$ and amplitude $A$ appear to diverge as $x_{o}$ approaches a critical value. (b) Elliptical Kepler orbits for the bound states are clearly seen when the measured $x$ and $\theta^{-}$ are displayed as radial and azimuthal coordinates respectively. (c) Trajectories in $x \,\mbox{-}\, \theta_{i}$ plane, $i = 1,2$, showing regions of bound and scattering trajectories. The grey curves are predicted by the far-field analysis: $1/x= 1/x_o + {\pi \over 8a}{\left(\cos 2\theta_{i} - \cos 2\theta_{io}\right)}$, where $\theta_{io}, \, x_{o}$ are the initial values. Red, blue and green represent restricted, bound and scattering regions respectively. (d) Amplitude vs minimum separation fits $1/(x_{o}^{-1} - x_{c}^{-1}) + c$ with $x_{c} = 1.02a$ and $c = 0.725a$ (red curve), qualitatively consistent with the asymptotic far-field prediction (blue) $1/({x_{o}^{-1} - \pi/4a})$. Alternatively, $x_{c}$ can be determined from the log-log plots of wavelength and amplitude vs $x_{c} - x_{o}$, giving $x_{c}= (1\pm 0.032)a$ (see Supplementary text). (e) Scaling of period $T$ with amplitude $A$, $T\sim A^{\nu}$, with $\nu \simeq 1.588 \pm 0.11$ consistent with the $3/2$ of Kepler's third Law.}
\end{center}
\end{figure}

Within the four-dimensional space of initial separations and orientations (Fig. \ref{fig:setup}), we begin with the symmetric case $\theta_{1}= 0 =\theta_{2}$, released at the same height. The resulting trajectories (Fig. \ref{fig:symmetric}) are symmetric, i.e. $\theta^+ \equiv \theta_{1}+\theta_{2}=0$ at all times. For small initial value $x_o$ of the horizontal separation $x$, the $\theta_i$ undergo full rotations and $x$ oscillates periodically [Fig. \ref{fig:symmetric}(a)], as noted in experiments and simulations by Jung \textit{et al.} \cite{shelley}. As $x_{o}$ is increased the wavelength and amplitude of the oscillations increase sharply [Fig. \ref{fig:symmetric}(a)], until the terminal motion seems to approach the linear trajectories of isolated Stokesian discs. Finite container height makes it impossible to establish experimentally the existence of a threshold value of $x_o$ at which the wavelength and amplitude actually diverge. A similar limitation applies to the numerical evidence for scattering orbits \cite{kim2} using an expansion in $a/R$ and the method of reflections \cite{HB1,kim1}. 

Working at leading order in $a/R$, we  construct an effective Hamiltonian approach to the disc-settling problem and map the symmetric case to the gravitational Kepler problem, thus establishing the transition between periodic and scattering orbits. We then go on to explain the behaviours seen in asymmetric settling. We begin with an isolated settling disc: the horizontal velocity of an isolated settling disc is $\dot{x_{1}} = F \alpha \sin 2\theta_{1}$, where $F$ is its buoyant weight and the mobility $\alpha$ is defined below. The tilt angle $\theta_{1}$ remains constant. We can thus view the trivial evolution of $x_{1}$ and $\theta_{1}$ as the Hamiltonian dynamics of a free particle with momentum $\theta_1$ and kinetic energy proportional to $\cos 2\theta_{1}$. This approach also applies to the two-disc case, where $\theta_1, \,\theta_2$ do not remain constant.  

For symmetric settling, retaining the lowest non-vanishing contribution in an expansion (see Supplementary Text) in $a/x$, $\dot{x} = 2F \alpha \sin \theta^{-}$ and $\dot{\theta^{-}} \equiv \dot{\theta_{2}} - \dot{\theta_{1}} = 2 F \gamma/x^{2}$. The proportionality constants $\alpha$ and $\gamma$ are determined by the solution for an isolated settling spheroid \cite{kim2,chwang}. The mobility $\alpha =- ({X_{A}}^{-1} - {Y_{A}}^{-1} )/{12\pi \mu a}$ and $\gamma =1/8\pi\mu$, where the resistance functions $X_{A} = 8/3\pi$ and $Y_{A} = 16/9\pi$ in the limiting case of $e = \sqrt{1 - b^{2}/a^{2}} \rightarrow 1$ for radius $a$ and thickness $b$ of the disc. The above far-field equations can be recast as Hamiltonian dynamics $\dot{x} = \partial_{\theta^{-}} \mathcal{H}, \,\dot{\theta^{-}} = -\partial_x \mathcal{H}$ with   
\begin{equation} 
\label{equation:1} 
\mathcal{H} \equiv 4F \alpha \ \sin^{2} {\theta^{-}\over{2}} + 2F \gamma /x
\end{equation} 

where $4F \alpha \ \sin^{2} {\theta^{-}\over{2}}$ and $2F \gamma / x$ play the roles of kinetic and potential energy respectively, with the $1/x$ coming from the viscous hydrodynamic kernel, not gravity. This is precisely the reduced Hamiltonian for the Kepler problem \cite{LLmech} when expressed in terms of azimuthal angle $\theta^-$ and radial coordinate $x$. The solution  
\begin{equation}
\label{equation:2}
   \frac{1}{x} - \frac{1}{x_{o}} = \frac{\alpha}{\gamma} (\cos \theta^{-} - \cos \theta^{-}_{o})
\end{equation}  
to the equations of motion, obtained earlier by Kim \cite{kim2} for far-field scattering trajectories, is simply conservation of $\mathcal{H}$, describing both bound and scattering orbits (see Fig. \ref{fig:symmetric}c), with a transition as $x_o \to x_{c}= 4a/\pi$. A circular Kepler orbit arises only for $\alpha=0$, which is the case of a pair of identical spheres. Given the very close approach of the discs in a bound state, the far-field mapping to the Kepler problem bears up surprisingly well against experimental observations, as detailed in Fig. \ref{fig:symmetric}.

 A simple case of asymmetric initial conditions  consists of releasing the discs at the same height with their normal vectors perpendicular to each other, $\theta^{+} = \pi/2$ [Figure 3 a-c]. Once again, periodic dynamics in the orientation is observed, with the added complexity of  $y$ oscillating between positive and negative values, and an apparent transition to unbounded orbits with increasing $x_{o}$.

\begin{figure}[b]
  \begin{center}
  \includegraphics[width=8.5cm]{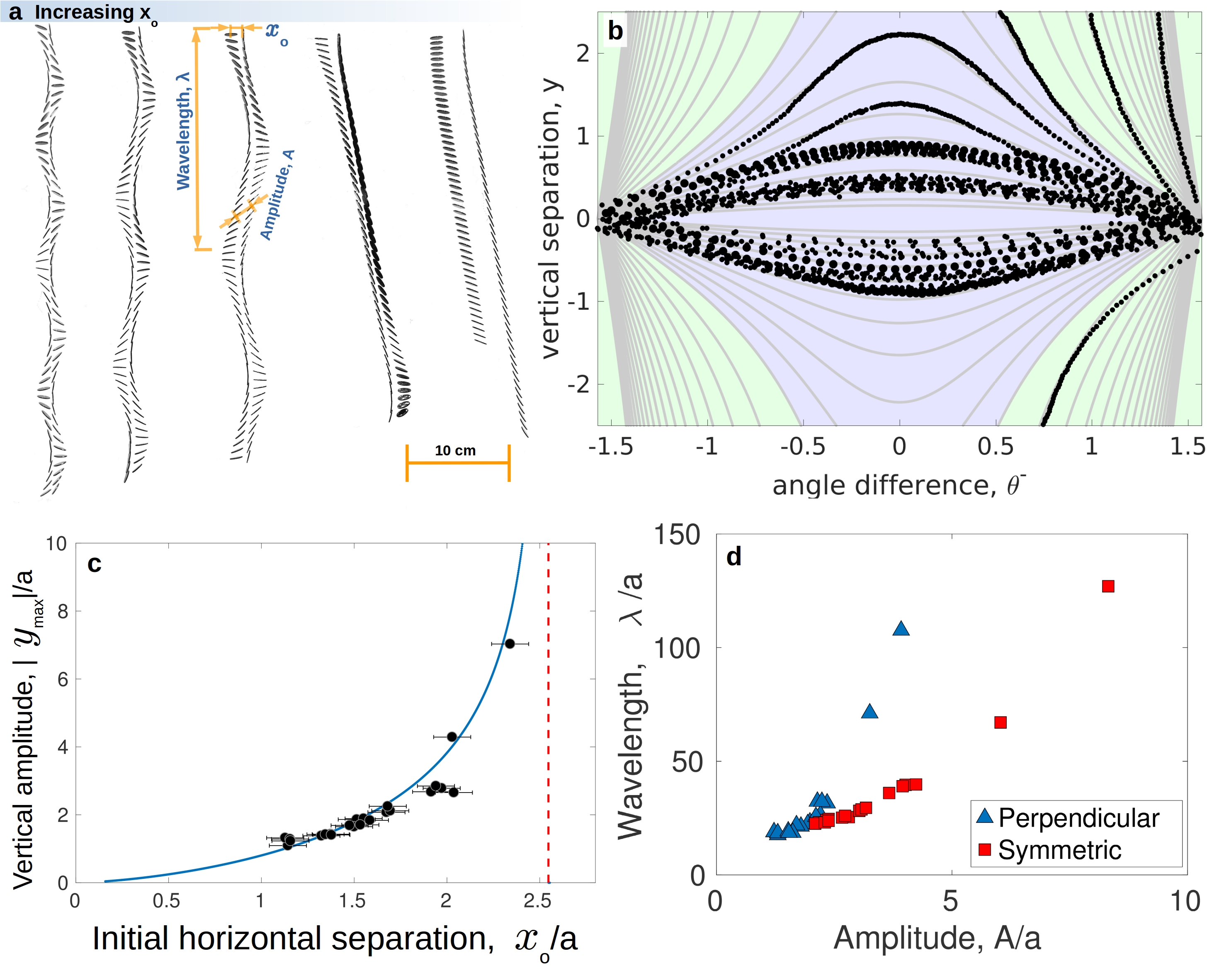}
  \caption{\label{fig:perpendicular}\textbf{Perpendicular initial condition}: (a) Overlapped time-lapse images from the experiment, when the discs are released with perpendicular initial orientation. As we move from left to right the initial horizontal separation $x_{o}$ is increased leading to a divergence in vertical separation $y$ (b) Experimental trajectories in the $\theta^{-}$-$y$ plane represented by points, compared with the far field result plotted in grey solid lines: $y = \pm \frac{x_{o} \cos \theta^{-}}{\sqrt{\left(8a/\pi x_{o}\right)^{2} - \cos ^{2} \theta^{-}}}$. Blue and green in the phase diagram represent bound and scattering regions respectively as predicted by far-field. (c) Divergence of amplitude of $y$ oscillations is captured by plotting the maximum value of $y/a$ as a function of initial horizontal separation $x_{o}/a$. The solid curve is the far-field prediction of amplitude: $A(x_{o}) = \frac{x_{o}}{\sqrt{\left(8a/\pi x_{o}\right)^{2} - 1}}$, with the red dotted line representing the critical $x_{o} = 8a/\pi$. (d) Observed wavelength $\lambda/a$ increases more strongly as a function of amplitude $A/a$ for perpendicular (blue) as compared to the symmetric case (red).} 
  \end{center}
  \end{figure}

\begin{figure}[h]
	\begin{center}
		\includegraphics[width=8.5cm]{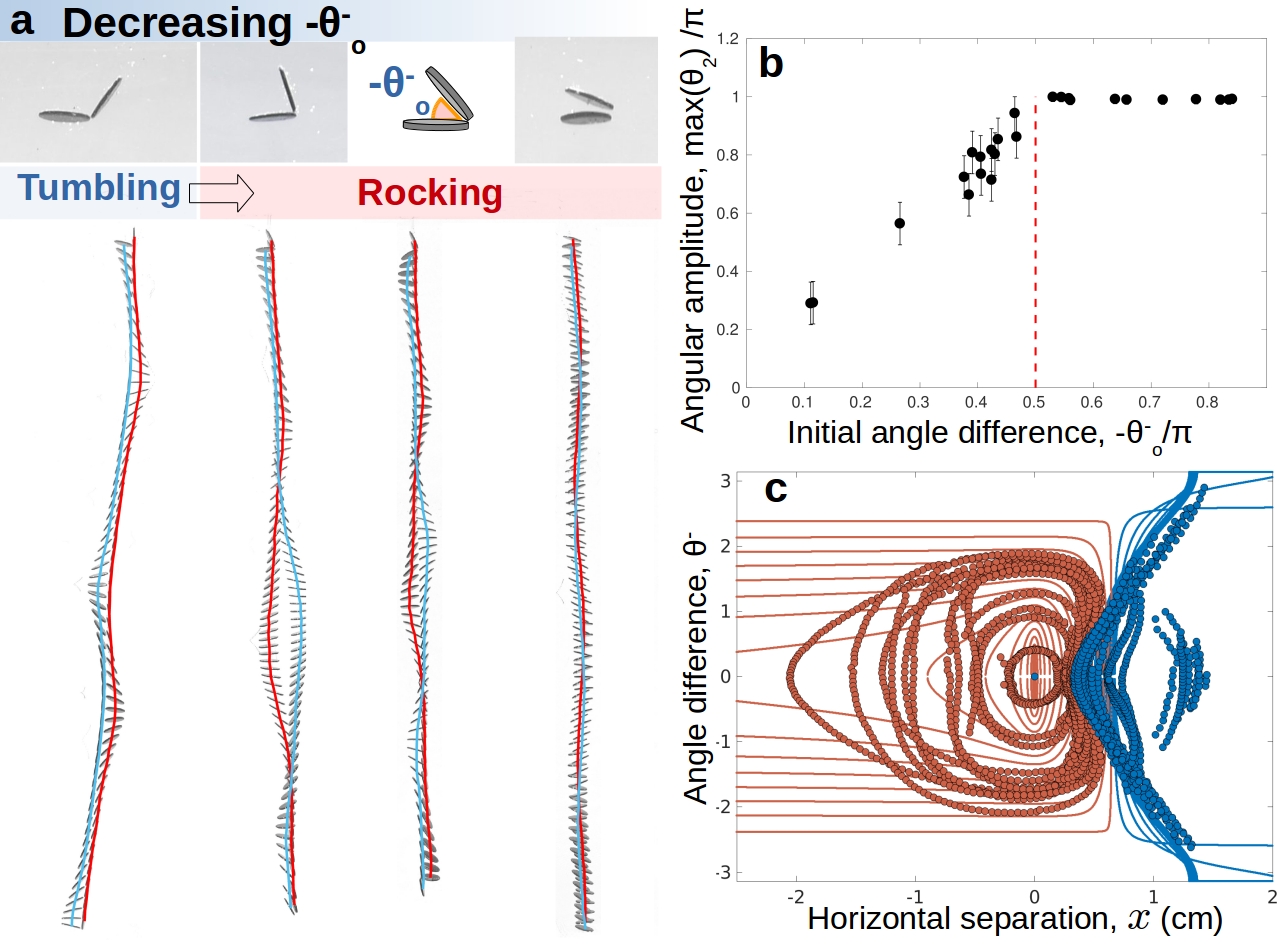}
		\caption{\label{fig:rocking}\textbf{Tumbling to rocking transition}: (a) As initial $\theta^{-}_{o}$ decreases there is a cross-over from tumbling to rocking dynamics. Trajectory of disc on the right (red) exchanges the relative $x$ position with the trajectory of disc in the left (blue) except for the first trajectory where $-\theta^{-}_{o}>\pi/2$. (b) To capture the transition from rocking to tumbling maximum angle of the disc at the right is plotted as a function of initial $-\theta^{-}_{o}$. We observe a transition from rocking to tumbling at initial $-\theta^{-}_{o} = \pi/2 $ (dotted red line), consistent with the far-field calculation. (c) The trajectories plotted in $x-\theta^{-}$ plane, red symbols represents rocking motion and blue represents tumbling. The corresponding red and blue solid curves represent the far-field prediction of rocking and tumbling dynamics respectively (see Supplementary). } 
	\end{center}
\end{figure}

The effective Hamiltonian description above provides a useful framework for understanding the dynamics resulting from a more general set of initial conditions $(x_{o},y_{o},\theta^{+},{\theta^{-}}_{o})$.
A reduction to an effective two-dimensional dynamics can be achieved for asymmetric initial conditions $\theta^+_o \neq 0$ as well, and periodic behaviour is preserved but more complex 
(Figs. \ref{fig:perpendicular} and \ref{fig:rocking}). The resulting non-Keplerian behaviour can be understood by extending equation \ref{equation:1} to incorporate the dependence of the angular velocity of the discs on the angle between the separation vector ${\bf R}$ and the external force ${\bf F}$. To leading order in $a/R$, the angular velocities of discs are equal and opposite, $\dot{\bf \theta_{1}} = - \dot{\bf \theta_{2}} = \gamma {\bf F} \times {\bf R} /R^{3}$. With this additional ingredient, we get the general equations of motion
\begin{align}
\dot{x} = 2 F \alpha \sin \theta^{-} \cos \theta^{+}, &&
\dot{y} = - 2 F \alpha \sin \theta^{-} \sin \theta^{+} \label{equation:3}\\
\dot{\theta^{-}} = 2 F \gamma\frac{x}{R^{3}}, && \dot{\theta^{+}} = 0. \label{equation:4}
\end{align}
Here $x \equiv x_{2} - x_{1}$, $y \equiv y_{2} - y_{1}$, $\theta^{-} \equiv \theta_{2} - \theta_{1}$ and $\theta^{+} \equiv \theta_{1} + \theta_{2}$ and $\alpha$ and $\gamma $ are defined above \eqref{equation:1}. The form \eqref{equation:3} and \eqref{equation:4} also follows on general grounds of symmetry (see Supplementary text). The conservation of $\theta^{+}$ in \eqref{equation:4} constrains the dynamics of $x$ and $y$ to a line with slope $-\tan\theta^{+}$, reducing the number of variables reduces to two, thus allowing phase plane analysis. The dynamics in terms of $S \equiv |{\bf R} - {\bf R}_o|$ and $\theta^{-}$ (see Supplementary text) is given by $\dot{S} = \partial_{\theta^{-}} \mathcal{H}, \,  
\dot{\theta^{-}} = - \partial_S\mathcal{H}$, with effective Hamiltonian
\begin{equation}
 \mathcal{H} \equiv  4F \alpha \ \sin^{2} {\theta^{-}\over{2}} + 2 F {\bar{\gamma}(S) \over R(S)}
\label{equation:5}
\end{equation}
 
where $\bar{\gamma}(S) \equiv \gamma \left( y_{o} - S \sin \theta^{+} \right)/ \left( y_{o}\cos \theta^{+} + x_{o} \sin \theta^{+}\right)$ and $R(S) = (S^2 + {R_{o}}^2 + 2S x_o \cos \theta^+ + 2S y_o \sin \theta^+)^{1/2}$. Note that $\theta^{+} \rightarrow 0$ yields the Keplerian limit for all initial separations $(x_{o},y_{o})$.

 The Hamiltonian \eqref{equation:5} for $\theta^{+} = \pi/2$ implies a dynamics with $y$ oscillating between positive and negative values, constant $x$, and, with increasing $x_{o}$, a transition from periodic to unbound orbits at $x_{c} = 8 a/\pi$ (see Supplementary text). These are in accord with observations (Fig. 3 and Supplementary Videos 3 \& 4), though the experiments additionally show small oscillations in $x$ possibly arising from near-field effects and small imprecision in initial release angles.  
For both symmetric and perpendicular initial conditions, the time period diverges at the boundary between bound and scattering orbits. Assuming the ratio of thickness to radius of discs is negligibly small, we have two length scales in the problem: the radius $a$ of the disc and the separation $R$ between the particles. One expects the period $T = ({a^{2} \mu}/{F}) f({R}/{a}, \text{Re}\,, \text{Fr}\,, \theta^{+}, {\theta^{-}}_{o})$, where the scaling function $f$ depends on the initial orientations, as well as on the Reynolds number Re $ = \rho U a/\mu \simeq 10^{-4}$  and Froude number Fr $= U/\sqrt{g a} \simeq 10^{-3}$ both of which are negligibly small. $f(R_{o}/a)$ can be calculated for symmetric and perpendicular cases (see Supplementary text) in the far-field limit, whence we find that the wavelength $\lambda \sim T F / a \mu$ diverges more strongly ($\sim A^{3}$) for the perpendicular case than for the symmetric case ($\sim A^{3/2}$ Kepler's 3rd Law), a trend consistent with our observations (see Figure 3d). 

Rocking -- a qualitatively distinct periodic behaviour analogous to libration in a pendulum, in which $\theta^{-}$ oscillates in a limited range -- emerges for $\pi/2 < \theta^{+} < \pi$. Releasing the discs with $\theta_{1} = \pi/2$ and decreasing $-\theta_{2}$ from $\pi/2$ (symmetric case) towards zero we experimentally capture the tumbling-rocking transition  at $\theta^{-}_o=-\pi/2$ (see Figure 4a and 4b). Unlike in tumbling, in rocking orbits the sign of $x$ and hence, from \eqref{equation:4}, of $\dot{\theta^{-}}$, alternates as the particles interchange their relative horizontal positions. Except for the special cases of parallel and perpendicular release, rocking dynamics is best viewed in $x$, $y$ and $\theta^{-}$ space albeit with proportional $x$ and $y$ displacements. Figure 4(c) shows the trajectories projected on the $x$-$\theta^{-}$ plane. The tumbling-rocking transition can once again be understood in terms of the effective Hamiltonian \eqref{equation:5} (see Supplementary text). 

Our experiments have uncovered a rich dynamics in the zero-Reynolds-number settling of a pair of identical discs, with a well-defined boundary between bound and scattering orbits and two distinct classes of periodic bound-state motion. Despite limited accuracy in locating the bound-scattering boundary, and excluding extreme situations where a disc is in the hydrodynamic shadow of another, the far-field hydrodynamic interaction offers a satisfactory and detailed understanding of the dynamics, even close to particle contact. Unexpectedly, the conservative dynamics generated by an effective Hamiltonian governs this viscosity-dominated system, with the tilt of the discs playing the role of momentum. For a large family of initial conditions the problem maps precisely to that of Kepler orbits. We also find and account for a distinct family of orbits with no planetary-orbit analogue, where the angle executes oscillations over a limited range. We expect our approach to offer insight into order and chaos in the settling dynamics of a wide class of anisotropic-particle systems.
 
\medskip

\noindent \textbf{METHODS}

The settling experiments were carried out in a quasi-two-dimensional container with dimensions of 30 cm x 50 cm x 5 cm (Width x Height x Depth). The viscous fluid was transparent polydimethylsiloxane (silicone oil) of viscosity 60000 cSt and density $0.96$ g cm$^{-3}$. The particles were made by punching out discs of diameter $d=1.3$ cm  from a 1mm thick aluminium sheet of density $2.7$ g cm$^{-3}$. They were made smooth using sandpaper and were spray painted black. Time-lapse images were captured every 5 seconds with a Nikon D7000 DSLR using the gphoto2 commandline interface. The imported images were converted to 8-bit, and thresholded after subtracting the background. The tracking was done by fitting an ellipse to the discs, with the centroid of the ellipse giving the positions ($x_{i}$, $y_{i}$) with an error of $\pm\,0.02 \,a$ and orientation of the major axis giving $\theta_{i}$ of the discs with an error of $\pm\, 0.061^{\circ}$.

{\bf Acknowledgements}: This work was begun when the authors were affiliated to the TIFR Centre for Interdisciplinary Sciences, Hyderabad. SR acknowledges support from the Tata Education and Development Trust and from a J C Bose Fellowship of the SERB, India. We thank R Govindarajan for discussions. The experimental work at UMass was supported through NSF DMR-1507650.

\pagebreak
\onecolumngrid
\widetext
\begin{center}
\textbf{\large Supplementary text}
\end{center}
\section{Error in ellipse fitting}
To find the error associated with fitting an ellipse in ImageJ, we track a single settling disc in the scattered state, where it follows a straight line path as shown in Figure 5 (a) and (b) . In Figure 5(c) residual with respect to linear fitting gives a measure of error. The measured root mean squared error is $0.012$cm or $0.019 a$ in terms of the radius $a$ of disc. Similarly Figure 5(d) shows the linear fit of angle of the major axis of the fitted ellipse with time giving a slope of -0.0006 and its residual in Figure 5(e) gives the root mean squared error of  0.061 degrees.  
\begin{figure*}[h]

\includegraphics[width=18cm]{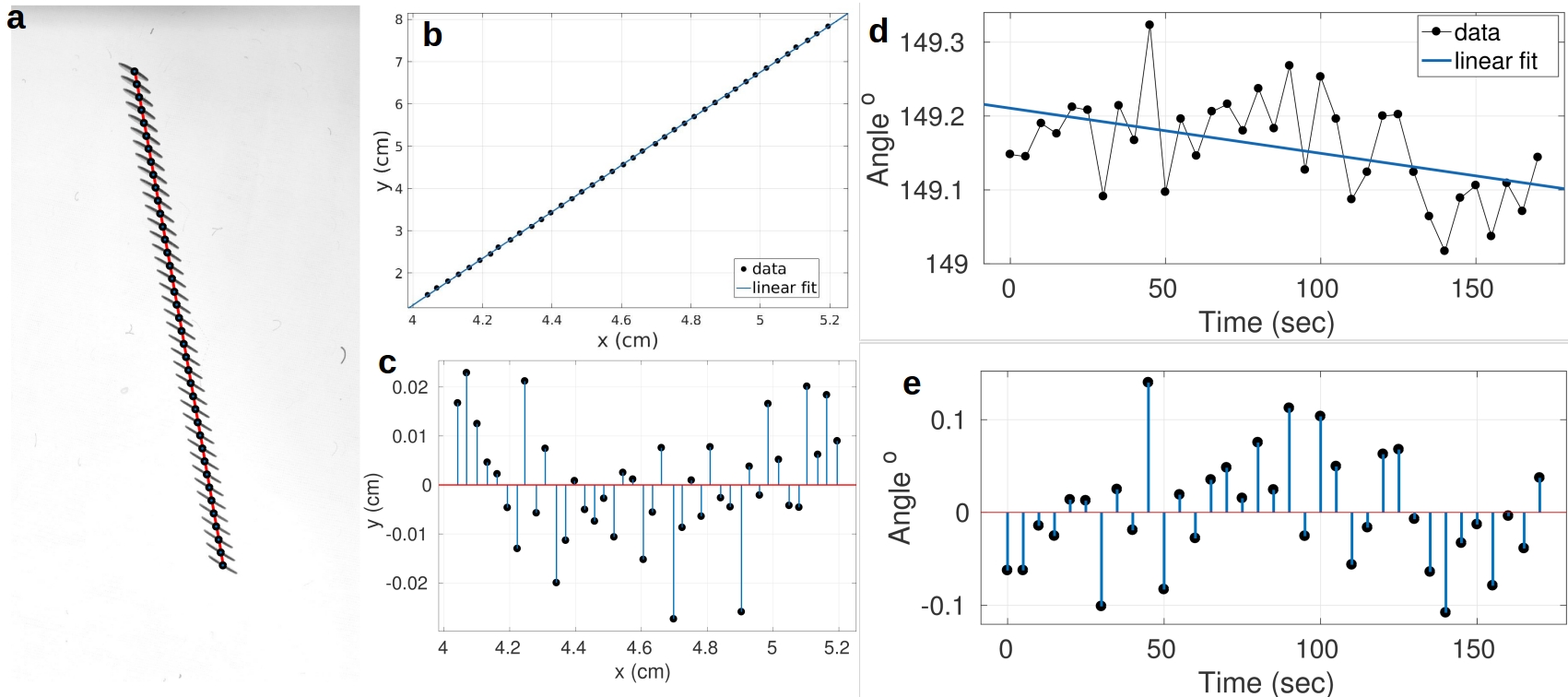}
	\caption{Systematic error. (a) the experimetnal z-stacks of a single disc superimposed with trajectory of the fitted ellipse (b) Linear fit of spatial trajectory of the centroid of fitted ellipse. (c) Residual plot of position (d) angle of the major axis of the fitted ellipse plotted against time. (e) Redidual plot of angle}
	\label{fig00}	
\end{figure*}
\section{log-log plot for symmetric divergence}
To determine the critical value of initial separation $d$ for which the wavelength $\lambda$, amplitude $A$ and time period $T$ diverges we plot $\log$ of $\lambda$,$A$ and $T$ as a function of $\log(x_{c} - d/0.65)$ with $x_{c}$ being a fitting parameter which simultaneously minimizes the difference between slope of individual curves. This minimisation gives $x_{c} = 1$ which is used for this plot. 
\begin{figure}[h]
	\includegraphics[width=8 truecm]{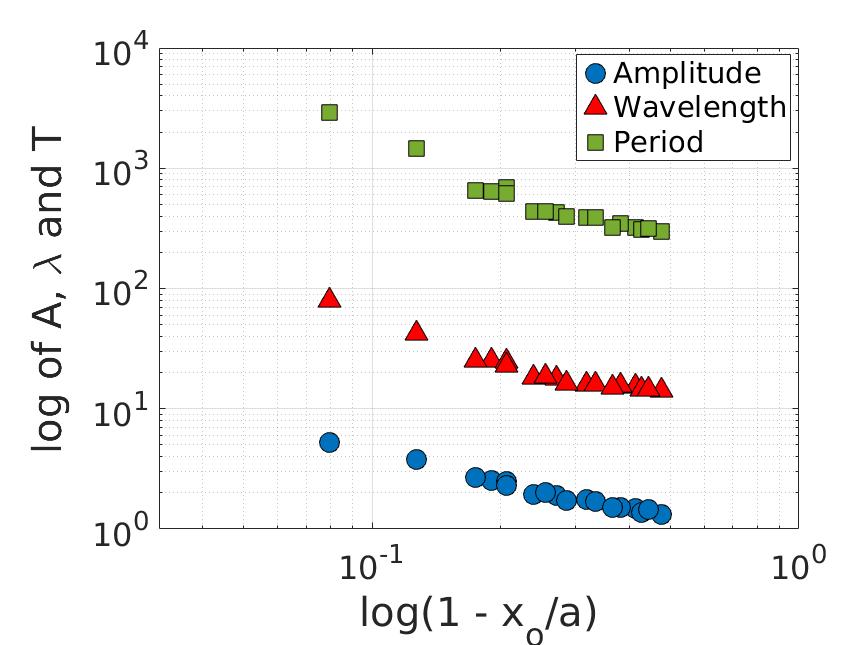}
	\caption{}
	\label{fig02}
\end{figure}
\\
\\
\\   
\section{Using symmetries}
\begin{figure}[b]
	\includegraphics[width=6 truecm]{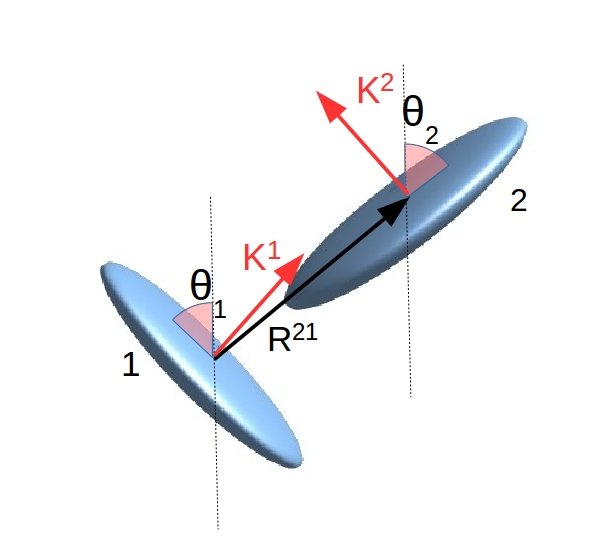}
	\caption{}
	\label{fig03}
\end{figure}
The system as shown is figure \ref{fig03} has three dynamical variables:
\begin{itemize}
	\item $\vec{K}^{1}$ , $\vec{K}^{2}$ , axisymmetric axis of the particles 1 and 2. 
	\item $\vec{R}^{12}$ separation vector pointing form particle 1 to 2. 
\end{itemize}
The equation of motion for the pair of particles is:
\begin{equation}
V_{i} = M_{ij}F^{j}
\label{eqn:01}
\end{equation}
where the left side is the time rate of change of the dynamical variables, $M_{ij}$ is the mobility tensor and $F^{j}$ is the external driving force directed vertically downwards. We construct a simple form of the second order tensor $M_{ij}$ allowed by the following symmetry of this system:
\begin{itemize}
	\item Time reversal symmetry, i.e. reversal of both $V$ and $F$.
	\item Apolarity of particles, $\vec{K} \longrightarrow - \vec{K}$, transform any one or both the particles.
	\item Particle exchange symmetry
	\item Rotational symmetry in the perpendicular subspace, $x \longrightarrow -x$ , $K_{x} \longrightarrow -K_{x}$ and $R_{x}^{12} \longrightarrow -R_{x}^{12}$.
\end{itemize}
Given the above symmetries the equation of motion for $R_{i}$ looks like
$$\frac{d R_{i}^{12}}{ dt} = M^{T}_{ij}F^{j}$$
The simplest form of $M^{T}_{ij}$ which satisfies the above symmetries is
$$M_{ij} = \alpha K_{i}^{1}K_{j}^{1} \, + \, \beta K_{i}^{2}K_{j}^{2} $$
where the coefficients $\alpha$ and $\beta$ can be arbitrary even scalar functions of $K_{l}R_{l}$. From the condition $\vec{R}^{12} = - \vec{R}^{21}$, we get the relation $\beta = -\alpha$. The spatial part of the dynamics is
\begin{equation}
\frac{d R_{x}^{12}}{dt} = |F| \alpha_{x}(K_{x}^{1}K_{y}^{1} \, - \, K_{x}^{2}K_{y}^{2}).
\label{eqn:02}
\end{equation}
 Substituting $K_{x}^{i} = \sin\theta_{i}$ and $K_{y}^{i} = \cos\theta_{i}$, where i = 1,2 is the particle index, gives
\begin{align}
\frac{d R_{x}^{12}}{dt} & = \frac{|F| \alpha_{x}}{2}(\sin 2\theta_{1} \, - \, \sin 2\theta_{2}) = |F|\alpha_{x} \, \sin(\theta_{1} - \theta_{2})\cos(\theta_{1} + \theta_{2}).
\label{eqn:03}
\end{align}
and similarly for the y component,
\begin{align}
\frac{d R_{y}^{12}}{dt} & = |F| \alpha_{y}(\cos^{2}\theta_{1} \, - \, \cos^{2}\theta_{2}) = |F|\alpha_{y} \, \sin(\theta_{2} - \theta_{1})\sin(\theta_{1} + \theta_{2}).
\label{eqn:04}
\end{align}
 Since the angular velocity is directed perpendicular to the plane of the dynamics, the simplest equation for the orientation angle allowed by the symmetries is
\begin{equation}
\left (\frac{d \theta_{1}}{ dt} \right )_{i} = \gamma \, \epsilon_{ijk} R_{j}^{12} F^{k}.
\label{eqn:05}
\end{equation}
Since $R^{21} = - R^{12}$ , we get $$\frac{d \theta_{1}}{ dt} = - \frac{d \theta_{2}}{ dt} $$
This condition is not true in general and breaks down for non-symmetric configurations if we allow rotational mobility to depend on $\vec{K}$. For example, term like $\epsilon_{ilm}K_{l}R_{m}R_{j} K_{k}R_{k} $ is allowed by symmetries but is not odd under $\vec{R} \longrightarrow -\vec{R}$. Also note that, the above prescription does not give us the exact form of the coefficients $\alpha_{x}$, $\alpha_{y}$ and $\gamma$ and determining their value requires hydrodynamic calculation.\\
\section{Dynamics from a far-field analysis}
The leading order translational response of a spheroid to an external force ${\bf F}$ is
\begin{equation}
 {\bf U_{1}}=  [{X_A}^{-1} {\bf K} {\bf K} + {Y_A}^{-1}({\bf \delta} - {\bf K} {\bf K})] \frac{{\bf F}}{6\pi \eta a}
 \label{eqn:06}
\end{equation}
Here ${\bf K}$ is the orientation vector, $X^{A}$ and $Y^{A}$ are resistance functions for spheroids \cite{KIM}. The second spheroid generates a velocity field which is calculated using the distribution of singularities at ${\bf r_{2}}$. For prolate case, it is a line distribution between the focal points:
\begin{equation}
{\bf v_{2}(r) = F_{2} \cdot \int_{-k_{2}}^{k_{2}} \left \{ 1 \, + \, (k_{2}^{2} - r_{2}^{2})\frac{(1-e_{2}^{2})^{2}}{4e_{2}^{2}} \nabla^{2} \right \} \frac{\mathcal{G}(r -r_{2})}{8\pi \eta} \, dr_{2} }
\label{eqn:07}
\end{equation}
Here $\mathcal{G}(r)_{ij}/8\pi \eta$ is the Green's function for Stokes flow and $k = a e$ with semi-major axis $a$ and eccentricity $e$. For the case of oblate spheroid the line integral is carried along a complex focal length $(k \longrightarrow ik)$, which is equivalent to a distribution of singularities on a disc of radius $k$ in the perpendicular real subspace \cite{Lisa}. Vorticity of ${\bf v_{2}}$ evaluated at the singularity distribution of the first particle, placed at $\bf{r_{1}}$, gives the leading contribution to the rotation of the first particle:
\begin{equation}
{\bf \omega_{1} = \frac{1}{8\pi \eta} \int_{-k_{1}}^{k_{1}} \frac{3}{4k_{1}^{3}} \, dr_{1} \, \int_{-k_{2}}^{k_{2}} \, \frac{ dr_{2}}{2k_{2}} \,(k_{1}^{2} - r_{1}^{2}) \frac{ F_{2} \times r_{12} }{|r_{12}|^{3}} }
\label{eqn:08}
\end{equation}
here, ${\bf r_{12} = r_{1} - r_{2}}$. When particles are far apart we can take the integrand $ {\bf F_{2} \times r_{12}} / r_{12}^{3}$ outside the integral and the remaining integral is unity. If $\bf{R}$ is pointing from the centroid of second particle to the centroid of first, the far-field approximation gives
\begin{equation}
{\bf \omega_{1}} = \frac{ {\bf F_{2} \times R} }{8\pi \eta R^{3}}
\label{eqn:09}
\end{equation}
\\
\textbf{\underline{Equations of motion}}\\
\\
Let the degrees of freedom of $i^{th}$ particle be $(x_{i},y_{i},\theta_{i})$, where i=1,2. This gives six coupled equations of motion:
\begin{align*}
\dot{x_{1}} = F\alpha \sin 2\theta_{1} &&
\dot{x_{2}} = F\alpha \sin 2\theta_{2}\\
\dot{y_{1}} = F\beta \cos^{2}\theta_{1}&&
\dot{y_{2}} = F\beta \cos^{2}\theta_{2}\\
\dot{\theta_{1}} = F\gamma \frac{(x_{1} - x_{2})}{R^{3}}&&
\dot{\theta_{2}} = F\gamma \frac{(x_{2} - x_{1})}{R^{3}}
\end{align*}

Here, $\alpha = - \frac{1}{12\pi a} \left [\frac{1}{X_A} - \frac{1}{Y_A} \right]$ , $\beta = - \frac{1}{6\pi a} \left [\frac{1}{X_A} - \frac{1}{Y_A} \right] = 2\alpha$ and $\gamma = \frac{1}{8\pi}$\\
\\
 Note that we have written these equations in the settling frame of the particles. In the relative coordinates: $x \equiv x_{2}-x_{1}$ , $ y \equiv y_{2} - y_{1}$, $\theta^{-} \equiv \theta_{2} - \theta_{1}$ and $\theta^{+} \equiv \theta_{2} + \theta_{1}$: 
\begin{align}
\dot{x} = 2 F \alpha \sin \theta^{-} \cos \theta^{+} &&
\dot{y} = - 2 F \alpha \sin \theta^{-} \sin \theta^{+} &&
\dot{\theta^{-}} = 2 F \gamma\frac{x}{R^{3}} && \dot{\theta^{+}} = 0
\label{eqn:10}
\end{align}
\section{Effective Hamiltonian}
\begin{figure}[h]
	\includegraphics[width=6 truecm]{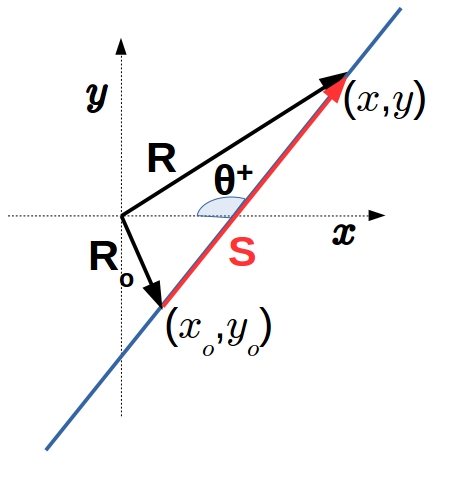}
	\caption{}
	\label{fig04}
\end{figure}

From equation \ref{eqn:10} we get the relation between $x$ and $y$
\begin{align}
\frac{d x}{d y} = - \frac{2\alpha}{\beta} \cot \theta^{+} = -\cot \theta^{+}; &&
x - x_{o} = -\cot \theta ^{+} (y - y_{o}).
\label{eqn:11}
\end{align}
Here, $x_{o}$ and $y_{o}$ are the initial $x$ and $y$ respectively. The conservation of $\theta^{+}$ can be used to rewrite equations \ref{eqn:10} in terms of the arc length $S$ along the line in the $x-y$ plane (see figure \ref{fig04}) and $\theta^{-}$.
$$S = |\vec{R}- \vec{R_{o}}| = \sqrt{(x - x_{o})^{2} + (y - y_{o})^{2}} = (x - x_{o}) \sec \theta^{+} $$
\begin{align}
\frac{d S}{d t} = \sec \theta^{+} \frac{d x}{d t} = 2F\alpha \sin \theta^{-} =  \frac{\partial  }{\partial \theta^{-}}(-  2F \alpha  \cos \theta^{-})
\label{eqn:12}
\end{align}
Using equation \ref{eqn:10} and the constraint imposed by equation \ref{eqn:11} we get  
\begin{align}
\frac{d \theta^{-}}{d t} & =  \frac{2 F \gamma x}{(x^{2} + \{y_{o} - \tan \theta^{+}(x - x_{o}) \}^{2})^{3/2}}  = -2 F \gamma \frac{d }{d x} \frac{y_{o} - \tan \theta^{+}(x-x_{o})}{(y_{o} + x_{o}\tan \theta^{+}) R}
\label{eqn:13}
\end{align}
Writing \ref{eqn:13} in terms of $S$ and $\theta^{-}$ gives
\begin{align}
 \frac{d \theta^{-}}{d t} = -2 F \gamma \sec\theta^{+} \frac{d }{d S} \frac{y_{o} - \tan \theta^{+}(x-x_{o})}{(y_{o} + x_{o}\tan \theta^{+})|\vec{S} + \vec{R_{o}}|} = - \frac{d }{d S} \frac{2F\gamma(y_{o} - S \sin \theta^{+})}{(y_{o}\cos \theta^{+} + x_{o}\sin \theta^{+})|\vec{S} + \vec{R_{o}}|}
 \label{eqn:14}
\end{align}

\section{Perpendicular settling}
\begin{figure}[h]
	\includegraphics[width=10 truecm]{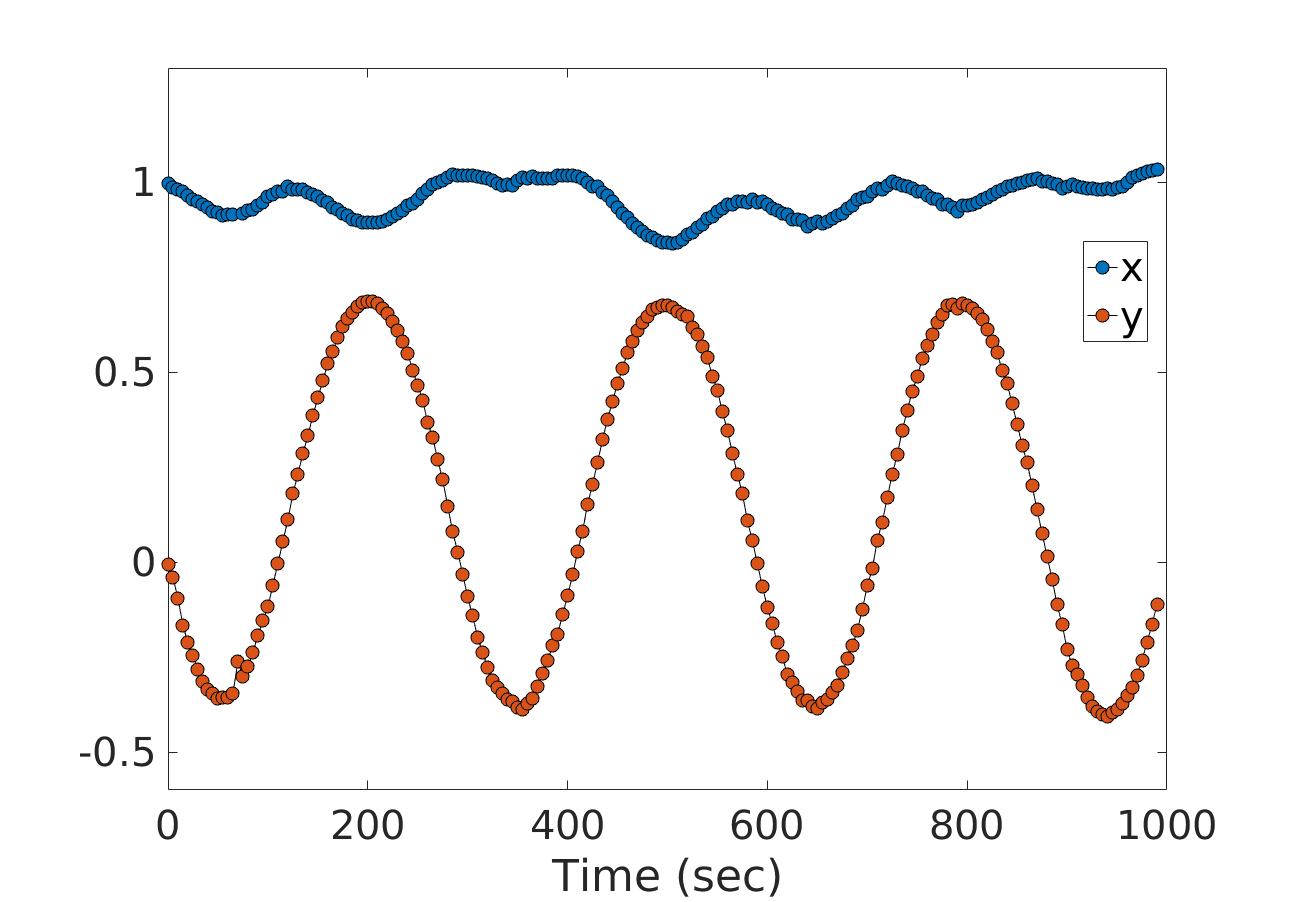}
	\caption{Observed vertical separation $y$ (blue) and horizontal separation $x$ (red) plotted against time, shows that $y$ oscillates between positive and negative values and $x$ is nearly constant}
	\label{fig05}
\end{figure}
Substituting the initial horizontal separation $y_{o} = 0 $ and $\theta^{+} = \pi/2$ in equation \ref{eqn:11}, \ref{eqn:12}, \ref{eqn:14} and solving for $x$ and $y$ in terms of $\theta^{-}$ gives
$$x = x_{o}.$$
and
$$\frac{\gamma x_{o} \, dy}{(x_{o}^{2} + y^{2})^{3/2}} = -\alpha \, \sin \theta^{-} \, d\theta^{-}$$
which upon integration and substituting the value $\alpha/\gamma = \pi/8a$ for disc, gives
\begin{align}
y = \pm \frac{ \cos \theta^{-} \, x_{o}}{\sqrt{\left(\frac{8a}{\pi x_{o}}\right)^{2} - \cos^{2} \theta^{-}}}.
\label{eqn:15}
\end{align}
This simplification when $\theta^{+} = \pi/2$ can be made use of for a wide range of initial conditions.

\section{Period of the orbits}
\textbf{(i)} $\bf{\theta^{+} \rightarrow 0}$\\
\\
For $\theta^{+} \rightarrow 0$ with arbitrary initial separation $R_{o} = \sqrt{{x_{o}}^{2} + {y_{o}}^{2}}$ and initial angle ${\theta^{-}}_{o}$, the resulting trajectory becomes ellipses if separation $R$ and angle difference $\theta^{-}$ is mapped to the radial coordinate and azimuthal angle respectively. 
\begin{align}
\frac{l}{R} =  1 + e\cos \theta^{-}
\label{eqn:16}
\end{align}
\begin{figure}[h]
	\includegraphics[width=8 truecm]{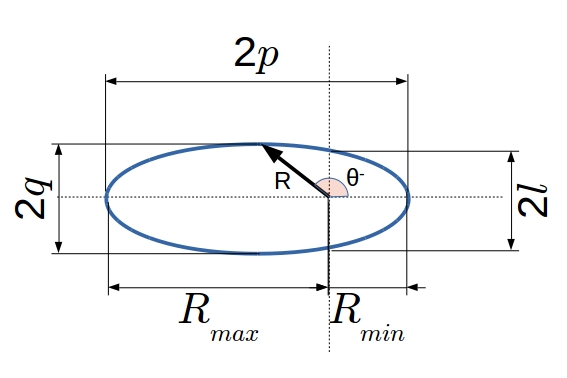}
	\caption{Bound orbit for symmetric case where $R$ and $\theta^{-}$ plays the role of radial and azimuthal coordinates respectively.}
	\label{fig06}
\end{figure}
where the \textit{latus rectum} $2l$, eccentricity of the orbit $e$, semi-major axis $p$ and semi-minor axis $q$ are
\begin{align*}
l = \frac{8a/\pi}{8a/ \pi R_{o} - \cos {\theta^{-}}_{o}} ; && e = \frac{1}{8a/\pi R_{o} -\cos {\theta^{-}}_{o}} ; && p = R_{max} R_{min} \left( \frac{1}{R_{o}} - \frac{\pi \cos {\theta^{-}}_{o}}{8a} \right); && q = (R_{max} R_{min})^{1/2} .
\end{align*}
here $R_{max} = \frac{8a/\pi}{8a/\pi R_{o} -  \cos {\theta^{-}}_{o} -1}$ is the maximum separation and $R_{max} = \frac{8a/\pi}{8a/\pi R_{o} -  \cos {\theta^{-}}_{o} + 1}$ is the minimum separation for a given initial separation $R_{o}$ and angle ${\theta^{-}}_{o}$. 
\subsubsection{Symmetric}
The trajectories becomes mirror symmetric when $y_{o} = 0$ making $R = x$ at all times. From equation \ref{eqn:13} we get $\dot{\theta^{-}} = 2F\gamma/x^{2}$ which is equivalent to the angular momentum conservation $R^{2}\dot{\theta^{-}} = 2F\gamma$. Using this conservation we can find the time period $T$ of the orbit by an integration which amounts to calculating area of the ellipse $\pi p q$.
\begin{align}
T = \frac{8 \pi^{2} \mu}{F} {R_{max}}^{3/2}{R_{min}}^{3/2} \left(\frac{\pi}{8a} + \frac{1}{x_{o}} - \frac{1}{x_{c}}\right)
\label{eqn:17}
\end{align}
here $x_{c} = 8a/\pi(\cos {\theta^{-}}_{o} + 1) $ is the critical value of initial $x_{o}$ at which the amplitude $R_{max}$ diverges. Time period can be written in the following form
\begin{align}
T = \frac{a^{2} \mu}{F} f\left(\frac{x_{o}}{a}, {\theta^{-}}_{o} \right)
\label{eqn:18}
\end{align}
where the scaling function
\begin{align}
f\left(\frac{x_{o}}{a}, {\theta^{-}}_{o} \right) = \frac{8^{3} \left( \frac{8a}{\pi x_{o}} - {\cos {\theta^{-}}_{o}} \right) }{\left( \frac{8a}{\pi x_{o}} - \cos {\theta^{-}}_{o} -1 \right)^{3/2} \left( \frac{8a}{\pi x_{o}} - \cos {\theta^{-}}_{o} + 1 \right)^{3/2}}
\label{eqn:19}
\end{align}
When $x_{o}$ approaches $x_{c}$ the time period diverges with amplitude as $T \sim A^{3/2}$. Since wavelength $\lambda$ scales as $\lambda \sim T F / a \mu$ it diverges the same way as $T$.\\
\\
\textbf{(ii)} $\bf{\theta^{+} \rightarrow \pi/2}$\\
\\
The solution of equation \ref{eqn:12} and \ref{eqn:14} in the limit $\theta^{+} \rightarrow \pi/2$ is
\begin{align}
\frac{y_{o}-y}{R} = \frac{\pi x_{o}}{8a} (\cos \theta^{-} - \cos {\theta^{-}}_{o})
\label{eqn:20}
\end{align}
\subsubsection{Perpendicular}
when the initial angle ${\theta^{-}}_{o} = \pi/2$ and initial $y_{o} =0$ the trajectories takes the form of ellipse: $l/ r =  1 + e\cos \phi$ if we map the square of separation $R^{2}$ to the radial coordinate $r$  and $2 \theta^{-} - \pi$ to the  azimuthal angle $\phi$. Here the \textit{latus rectum} $2l$, eccentricity of the orbit $e$, semi-major axis $p$ and semi-minor axis $q$ are
\begin{align*}
l = \frac{2 a^{2}}{2 a^{2}/ {x_{o}}^{2} - (\pi/8)^{2}} ; && e = \frac{(\pi/8)^{2}}{2 a^{2}/ {x_{o}}^{2} - (\pi/8)^{2}} ; && p = \frac{{R_{max}}^{2}}{2} {\left( 2(8a/\pi x_{o})^{2} - 1 \right)}; && q = \frac{8a}{\pi}R_{max} .
\end{align*}
here $R_{max} = {x_{o}}/{\sqrt{(8a/\pi x_{o})^{2} -1}}$ is the maximum separation or amplitude of the oscillation. From equation \ref{eqn:14} we have
\begin{align}
R^{3}\frac{d\theta^{-}}{dt} = 2F\gamma x_{o}
\label{eqn:21}
\end{align}
rewriting it in terms of $r$ and $\phi$
\begin{align}
\frac{1}{2\sqrt{r}}r^{2}{d\theta^{-}} = 2F\gamma x_{o} {dt}
\label{eqn:22}
\end{align}
integrating the left hand side by parts
\begin{align}
\frac{1}{\sqrt{r}} \int \frac{r^{2}}{2}d\phi \quad  -\int \frac{\sqrt{r}}{8} \left(\frac{\pi}{8a}\right)^{2} \, \sin \phi \, d\phi \, \int \frac{r^{2}}{2}d\phi \, = \, 2F\gamma x_{o} T
\label{eqn:23}
\end{align}
over one complete cycle $\phi$ goes from $0$ to $2\pi$ . The indefinite integral over $r^{2}d\phi/2$ is the area of sector of an ellipse which does not simplify $\ref{eqn:23}$ any further. To get a closed form of period as a function of $x_{o}$ we approximate the area of an eliptical sector by its supremum value $p \, q\,\phi/2$ . The second integral on the left hand side can be written completely in terms of $\phi$ using equation \ref{eqn:20}
\begin{align}
\frac{\pi \, p \, q}{x_{o}} \quad  -\,\frac{p\,q}{16}\,\int_{0}^{2\pi} \left(\frac{\pi}{8a}\right) \,\frac{ \phi \,\sin \phi}{\sqrt{\left(\frac{8a}{\pi x_{o}}\right)^{2} - \frac{1}{2} + \frac{\cos \phi}{2}}} \, d\phi \, \, = \, 2F\gamma x_{o} T
\label{eqn:24}
\end{align}
integrating the left hand side again by parts gives
\begin{align}
\frac{\pi \, p \, q}{x_{o}} \quad  -\,\frac{\pi\,p\,q}{32 a}\,\int_{0}^{2\pi} \, d\phi \, \,{\sqrt{\left(\frac{8a}{\pi x_{o}}\right)^{2} - \frac{1}{2} + \frac{\cos \phi}{2}}}  \, = \, 2F\gamma x_{o} T
\label{eqn:25}
\end{align}
the above integral has the form of complete elliptic integral $E[\phi/2 , (\pi x_{o}/8a)^{2}]$. This can be solved for time period $T$ in terms of $x_{o}$ by inseting the value for $p$ and $q$
\begin{align}
  T = \frac{32\, a\,{x_{o}} \, \mu}{F} \, f(x_{o}/a)
  \label{eqn:26}
\end{align}
here the scaling function $f(x_{o}/a)$ is
\begin{align}
f(x_{o}/a) = \,\frac{2 \left(\frac{x_{c}}{x_{o}}\right)^{2} -1 }{\left[ \left(\frac{x_{c}}{x_{o}}\right)^{2} - 1 \right]^{3/2}} \, \left( \pi \, - \, \left(\frac{x_{o}}{x_{c}}\right)^{2} \, E\left[(x_{o}/x_{c})^{2} \right] \right)
\label{eqn:27}
\end{align}
The critical value of $x_{o}$ at which $f(x_{o}/a)$ diverges is $x_{c} = 8a/\pi$. As $x_{o} \rightarrow x_{c}$ the time period $T$ diverges as
\begin{equation}
 T = \frac{256\, a^{2}\, \mu}{F{\left[ \left(\frac{x_{c}}{x_{o}}\right)^{2} - 1 \right]^{3/2}}} \,
 \label{eqn:28}
\end{equation}
in terms of amplitude $A$, time period goes as $T \sim A^{3}$ as $x_{o} \rightarrow x_{c}$.
\begin{figure}[h]
	\includegraphics[width=9 truecm]{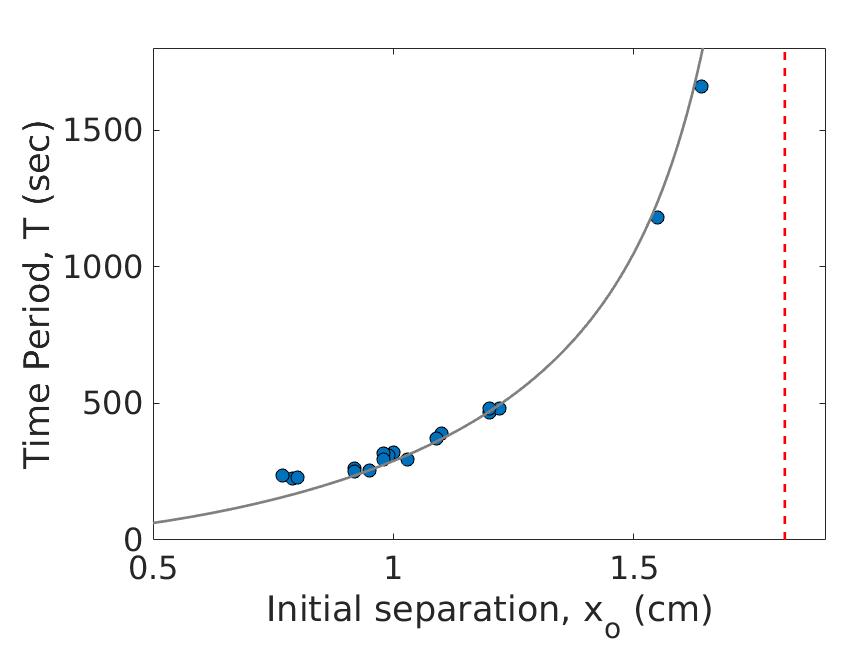}
	\caption{The data plotted along with the far-field form of $T$ with $a\mu/F$ = 2.2 and $x_{c} = 1.85$.}
	\label{}
\end{figure}
\section{Rocking trajectories}
\begin{figure}[t]
	\includegraphics[width=10 truecm]{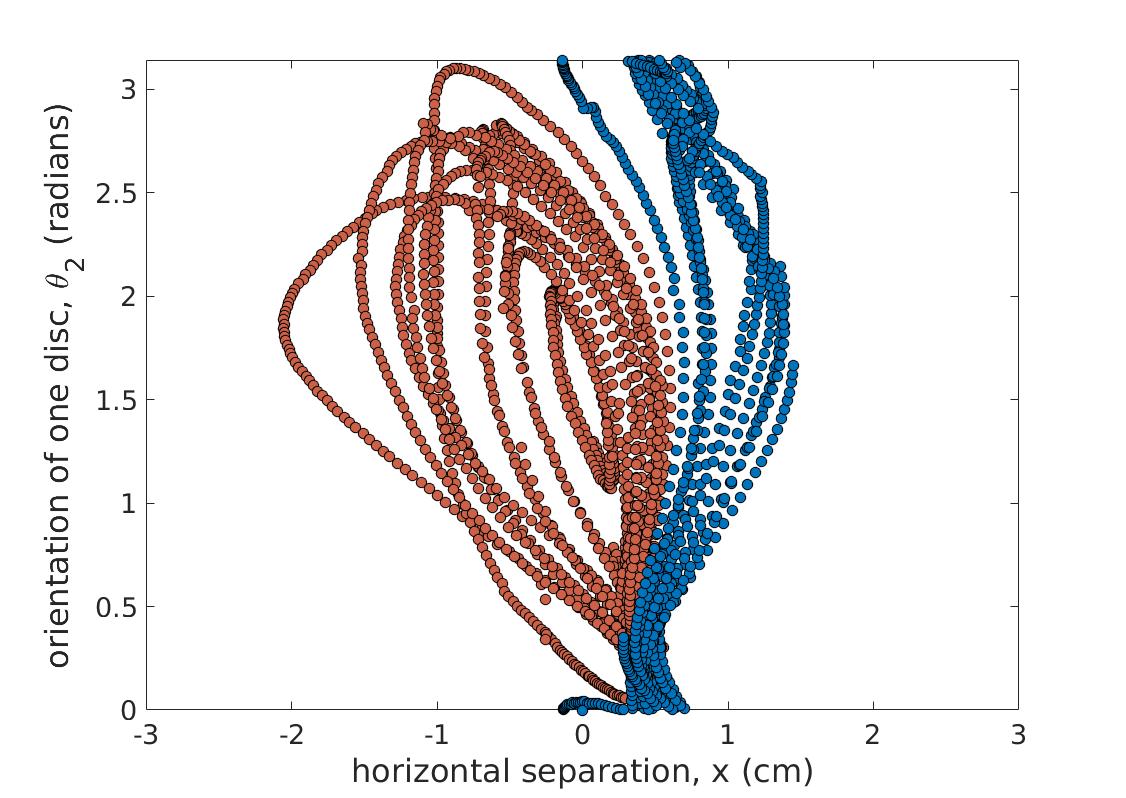}
	\caption{}
	\label{fig07}
\end{figure}

For initial $\theta^{-}_{o}<\pi/2$ both the disc rotates in a range of angles which defines the amplitude of angular oscillation. figure \ref{fig07} shows the trajectories of one of the disc for $\theta^{-}_{o}<\pi/2$ which exhibits oscillation in a range of angles (red) and $\theta^{-}_{o}>\pi/2$ showing tumbling orbits (blue).
Using equation \ref{eqn:10} we construct the dynamics in $x-\theta^{-}$ plane:
\begin{align*}
\frac{x dx}{(x^{2} + y^{2})^{3/2}} = \frac{\pi \cos \theta^{+}}{8 a} \sin \theta^{-}  d\theta^{-}
\end{align*}
using the constraint of equation \ref{eqn:11}
\begin{align}
\frac{x dx}{(x^{2} + \{y_{o} - \tan \theta^{+}(x - x_{o}) \}^{2})^{3/2}} = \frac{\pi \cos \theta^{+}}{8 a} \sin \theta^{-}  d\theta^{-}
\label{eqn:29}
\end{align}
defining $I \equiv x_{o}\tan \theta^{+} + y_{o}$ we can rewrite equation \ref{eqn:29}:
\begin{align}
\frac{x dx}{(x^{2}\sec^{2}\theta^{+} - 2I\tan \theta^{+} x + I^{2})^{3/2}} = \frac{\pi \cos \theta^{+}}{8 a} \sin \theta^{-}  d\theta^{-}
\label{eqn:30}
\end{align}
which upon integration gives:
\begin{align}
\frac{y_{o} - \tan \theta^{+}(x-x_{o})}{I\sqrt{x^{2}\sec^{2}\theta^{+} - 2I\tan \theta^{+} x + I^{2}}} - \frac{y_{o}}{I\sqrt{x_{o}^{2} + y_{o}^{2}}} = \frac{\pi \cos \theta^{+}}{8 a} (\cos \theta^{-} - \cos{\theta^{-}_{o}} )
\label{eqn:31}
\end{align}
which can be solved for $\theta^{-}$ in terms of $x$:
\begin{align}
\theta = \pm \cos^{-1} \left[ \frac{8 a}{I\pi \cos \theta^{+}} \left\{ \frac{y_{o} - \tan \theta^{+}(x-x_{o})}{\sqrt{x^{2}\sec^{2}\theta^{+} - 2I\tan \theta^{+} x + I^{2}}} - \frac{y_{o}}{\sqrt{x_{o}^{2} + y_{o}^{2}}} \right\} + \cos{\theta^{-}_{o}} \right]
\label{eqn:32}
\end{align}

\end{document}